\documentclass[runningheads]{llncs}
\usepackage[table]{xcolor}
\usepackage{booktabs}
\usepackage[table]{xcolor}

\definecolor{bestgreen}{RGB}{103,189,99}
\definecolor{midgreen}{RGB}{190,226,144}
\definecolor{loworange}{RGB}{247,150,30}
\definecolor{midorange}{RGB}{247,183,84}
\usepackage{booktabs}
\usepackage[T1]{fontenc}
\usepackage{graphicx}
\usepackage{amsmath}
\usepackage{amssymb}
\usepackage{booktabs}
\usepackage{url}
\usepackage{hyperref}
\begin{document}
\title{Pre- to Post-Contrast Synthesis of Breast DCE-MRI using Latent Bridge Matching}
\titlerunning{Breast DCE-MRI Synthesis with Latent Bridge Matching}
\author{Sina Amirrajab\inst{1} 
Zohaib Sallahuddin\inst{1}  Henry C Woodruff \inst{1,2}, \and Philippe Lambin \inst{1,2}}

\authorrunning{S. Amirrajab et al.}
%
\institute{The D-Lab, Department of Precision Medicine, GROW -- Research Institute for Oncology and Reproduction, Maastricht University, Maastricht, the Netherlands \and
Department of Radiology and Nuclear Medicine, GROW - Research Institute for Oncology and Reproduction, Maastricht University, Medical Center+, Maastricht, The Netherlands
}
\maketitle

\begin{abstract}
Dynamic contrast-enhanced magnetic resonance imaging (DCE-MRI) is central to breast cancer imaging, but gadolinium administration increases scan burden and motivates contrast-reduced alternatives, including synthetic contrast generation. We propose a latent bridge matching (LBM) framework for synthesizing peak-enhanced breast DCE-MRI from pre-contrast images in the MAMA-SYNTH challenge setting. Instead of starting from Gaussian noise as in conventional latent diffusion models (LDMs), the proposed model learns a conditional bridge between paired pre-contrast and peak-enhanced VAE latents. A latent UNet predicts the remaining correction from intermediate bridge states to the peak-enhanced latent, enabling iterative refinement while keeping the trajectory anchored to patient-specific anatomy. We evaluated two LBM conditioning variants on 91 DUKE validation cases. For the tumor-conditioned variant, tumor masks were used as conditioning inputs. Tumor-conditioning improved performance compared with pre-contrast conditioning, reducing MSE from 1.023 to 0.940 and FRD from 7.523 to 4.716, while increasing tumor SSIM from 0.355 to 0.429. The tumor-conditioned LBM also outperformed the evaluated LDM baseline on this validation cohort. These results suggest that latent bridge matching is a promising pre-contrast-anchored formulation for virtual contrast enhancement, while further work is needed to validate generalization and remove dependence on ground-truth tumor masks at inference. 
\end{abstract}
{\let\thefootnote\relax\footnotetext{The code is available at \url{https://github.com/sinaamirrajab/mama-synth-lbm}}}

Dynamic contrast-enhanced MRI (DCE-MRI) is a key component of breast cancer imaging because post-contrast enhancement reflects vascular and permeability-related tissue properties. However, gadolinium-based contrast administration increases protocol complexity, patient burden, and cost, and repeated exposure remains a concern in longitudinal imaging~\cite{Iyad2023GadoliniumReview}. Virtual contrast enhancement aims to synthesize post-contrast information from pre-contrast image, potentially reducing contrast use while approximating clinically relevant enhancement patterns.

The MAMA-SYNTH challenge provides a standardized benchmark for this problem in breast MRI~\cite{mamasynth2026}. The task is to synthesize a two-dimensional peak-enhanced post-contrast slice from the corresponding pre-contrast DCE-MRI input, with cases derived from the multi-center MAMA-MIA cohort~\cite{Garrucho2025MAMAMIA}. The challenge input is provided in a training-set z-score domain, and methods are evaluated with complementary metrics that include pixel-level error, perceptual similarity, tumor-region structural similarity, radiomic realism, and downstream task performance. This makes the task both an image-to-image synthesis problem and an intensity-calibration problem.

Existing breast MRI virtual contrast approaches have mainly relied on direct image translation, conditional GANs, U-Net-based synthesis, or diffusion-based generation~\cite{Osuala2025DynamicTumorCGAN,Osuala2024ContrastKineticsLDM,Osuala2024PrePostSegmentation}. These methods have shown that synthetic contrast-enhanced breast MRI is feasible, but they do not explicitly model the synthesis trajectory as a bridge between the paired pre-contrast and peak-enhanced representations.

We formulate peak-enhanced DCE-MRI synthesis as paired latent transport, where the enhancement map is learned by transforming the pre-contrast to peak-enhancement in latent space. Our method, based on the concept of latent bridge matching (LBM)~\cite{chadebec2025lbmlatentbridgematching}, encodes pre-contrast and peak-enhanced images into a VAE latent space and trains a latent UNet to predict the remaining correction from intermediate bridge states to the peak-enhanced latent. Unlike standard latent diffusion models (LDM)~\cite{Rombach2022LatentDiffusion}, LBM does not synthesize from pure noise; it starts from the observed source representation and iteratively transports it toward the target representation. Compared with residual regression, the bridge formulation exposes the network to intermediate off-source states during training and supports iterative refinement, while remaining more directly anchored to the input anatomy than conventional conditional diffusion.

Our main contributions are: (i) we formulate peak-enhanced breast DCE-MRI synthesis as a paired latent bridge matching problem; (ii) we condition the bridge on the pre-contrast latent representation, with optional tumor-mask guidance, to preserve patient-specific anatomy while modelling localized enhancement; and (iii) we evaluate the approach in the MAMA-SYNTH setting using complementary pixel-level, perceptual, and tumor-region radiomic metrics.

\section{Method}
\label{sec:method}

\begin{figure}[t]
    \centering
    \includegraphics[width=\linewidth]{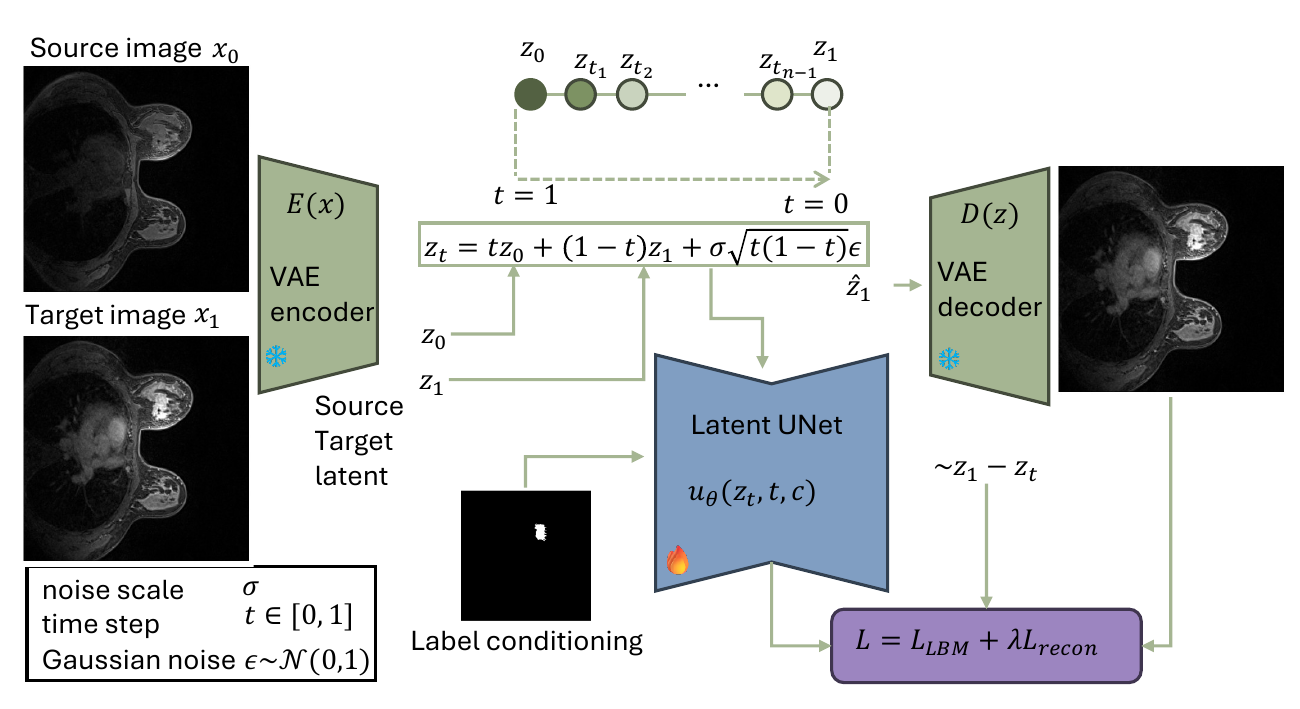}
   \caption{Overview of latent bridge matching for breast DCE-MRI synthesis. The pre-contrast image $x_0$ and peak-enhanced target image $x_1$ are encoded into source and target latents $z_0$ and $z_1$. During training, noisy bridge states $z_t$ are sampled between the endpoints using bridge time $t$, with $t=1$ at the source and $t=0$ at the target. A latent UNet predicts the remaining correction $z_1-z_t$, optionally conditioned on a tumor mask. At inference, the source latent is iteratively refined along decreasing bridge times and decoded to synthesize the peak-enhanced image.}
    \label{fig:method}
\end{figure}

\subsection{Problem Formulation}

Figure \ref{fig:method} shows an overview of the method. Let $x_0 \in \mathbb{R}^{H\times W}$ denote the pre-contrast source DCE-MRI slice and $x_1 \in \mathbb{R}^{H\times W}$ denote the corresponding peak-enhanced post-contrast target slice. Both images are represented in the challenge z-score domain. Before VAE encoding, intensities are mapped to the image range expected by the Stable Diffusion VAE. For lower and upper z-score bounds $a$ and $b$, we use
\begin{equation}
    x_{01} = \mathrm{clip}\left(\frac{x_z-a}{b-a},0,1\right),
    \qquad
    x_{\mathrm{VAE}} = 2x_{01}-1 .
\end{equation}
A VAE encoder $E$ maps source and target images to latent variables,
\begin{equation}
    z_0 = sE(x_0), \qquad z_1 = sE(x_1),
\end{equation}
where $s$ is a dataset-specific latent scaling factor. The synthesis objective is to estimate the target latent $z_1$ from the source image, while preserving patient anatomy and producing contrast-dependent intensity changes.

\subsection{Latent Bridge Matching}

For a bridge time $t\in[0,1]$, where $t=1$ corresponds to the source latent and $t=0$ corresponds to the target latent, we sample an intermediate bridge state
\begin{equation}
    z_t = t z_0 + (1-t)z_1 + \sigma\sqrt{t(1-t)}\,\epsilon,
    \qquad \epsilon\sim\mathcal{N}(0,I),
    \label{eq:bridge}
\end{equation}
where $\sigma$ is the bridge noise scale. Equivalently, the time-dependent noise level is $\sigma(t)=\sigma\sqrt{t(1-t)}$, which vanishes at both endpoints and is largest near the midpoint of the bridge.

The latent UNet $u_\theta$ is trained to predict the remaining correction from the current bridge state to the peak-enhanced target latent,
\begin{equation}
    u_\theta(z_t,t,c) \approx z_1-z_t,
    \label{eq:remaining_target}
\end{equation}
where $c$ denotes conditioning information. The training objective is
\begin{equation}
    \mathcal{L}_{\mathrm{LBM}} =
    \mathbb{E}_{(z_0,z_1),t,\epsilon}
    \left[
    \rho\left(u_\theta(z_t,t,c) - (z_1-z_t)\right)
    \right],
    \label{eq:lbm_loss}
\end{equation}
with $\rho$ implemented as an equally weighted combination of latent $\ell_1$ and $\ell_2$ terms. This state-dependent target matches iterative inference: at each point on the trajectory, the network predicts the correction that remains toward the peak-enhanced latent.



At inference, only the pre-contrast image is available. We encode it as
$\hat{z}^{(0)}=sE(x_0)$ and refine it over a decreasing bridge-time grid
$t_k=(K-k)/K$, for $k=0,\ldots,K$, with $t_0=1$ and $t_K=0$.
Since the network predicts the remaining correction to the target latent, we
use the relaxation coefficient
\begin{equation}
    \alpha_k = \frac{t_k-t_{k+1}}{t_k},
    \qquad k=0,\ldots,K-1 .
\end{equation}
The latent update is
\begin{equation}
    \hat{z}^{(k+1)}
    =
    \hat{z}^{(k)}
    +
    \alpha_k
    u_{\theta}\left(\hat{z}^{(k)},t_k,c\right).
    \label{eq:inference}
\end{equation}
After refinement, the final predicted target latent is decoded with the VAE
decoder:
\begin{equation}
    \hat{x}_1 = D\left(\hat{z}^{(K)}\right).
\end{equation}

\subsection{Relation to Latent Diffusion and Flow Matching}

Latent diffusion models learn to reverse a noise corruption process in latent space and are effective for broad image generation~\cite{Rombach2022LatentDiffusion}. In paired medical image translation, however, the source image already supplies patient-specific anatomy and most spatial structure. Starting the synthesis trajectory from random noise is therefore unnecessary and can make faithful pixel-level translation more difficult. LBM instead learns transport between paired source and target latents. This connects the method to flow matching and rectified flow, which learn vector fields along paths between distributions~\cite{Lipman2023FlowMatching,Liu2023RectifiedFlow}, while specializing the path to patient-conditioned source-to-target synthesis.

\subsection{Conditioning}

We evaluate two conditioning regimes. In the source-only model, the UNet receives the current bridge latent concatenated with the pre-contrast source latent. In the segmentation-conditioned model, a tumor mask is additionally encoded to the latent space and concatenated with the source latent representation. This gives explicit tumor-region information to the vector field, while leaving the bridge state responsible for synthesizing the target image. In the validation experiments reported here, the segmentation-conditioned model uses the validation tumor masks available for the validation set; at submission time, these masks are replaced by predictions from an auxiliary pre-contrast segmentation model.

\subsection{Implementation Details}

For each subject, we used the tumor-centered slice with maximum tumor area and included the two neighboring slices above and below it, yielding a five-slice window around the largest tumor extent. The source, target and tumor-mask slices were resized to $512\times512$. To make LBM training efficient, all source, target, and conditioning latents were computed once with the VAE and saved before training, so that the bridge model could be trained directly in latent space without repeatedly running the encoder.

All experiments use the Stable Diffusion VAE fine-tuned for MSE reconstruction and a latent UNet initialized from Stable Diffusion v1.5. Latents are extracted from the VAE posterior mode. Because breast MRI intensities and latent statistics differ from natural images, we rescale latents by the inverse standard deviation measured over pooled source and target training latents, giving $s\approx 0.128$. This replaces the default Stable Diffusion latent scale of $0.18215$.

The selected models use an adaptive peak-intensity normalization before VAE encoding. The input images are already in the challenge z-score domain. A small source-only histogram MLP predicts an upper value for min-max normalization for the expected peak-enhanced target from the pre-contrast slice. The predicted value is then used to normalize the pre-contrast image to the range of $[0,1]$ and de-normalize the synthesis post-contrast to the z-score range.

The bridge model is trained to map directly from pre-contrast latents to peak-enhanced latents using the remaining-correction objective. During training, bridge times $t$ are sampled uniformly from 500 evenly spaced values in $[0,1]$; all time values are not used for every image pair in every iteration. The bridge noise scale is set to $\sigma=0.005$. The latent loss is the average of $L_1$ and $L_2$ norms. No image-space reconstruction, tumor-image, enhancement, or tumor-enhancement losses are used for the reported models. Optimization uses AdamW with learning rate $9\times10^{-5}$, a constant learning rate schedule, 200 warmup steps, and mixed-precision training. Models are trained for up to 8000 steps. At validation time, synthesis uses 500 iterative bridge refinement steps.

\begin{table}[t]
\centering
\caption{Validation performance on 91 DUKE validation cases. Lower is better for MSE, LPIPS, and FRD; higher is better for tumor SSIM. Darker green indicates better performance.}
\label{tab:validation}
\small
\setlength{\tabcolsep}{2pt}
\renewcommand{\arraystretch}{1}
\begin{tabular}{lcccc}
\toprule
Model 
& MSE $\downarrow$ 
& LPIPS $\downarrow$ 
& Tumor SSIM $\uparrow$ 
& FRD $\downarrow$ \\
\midrule

LBM, source only 
& \cellcolor{midorange} 1.023 $\pm$ 1.169
& \cellcolor{midorange} 0.119 $\pm$ 0.035
& \cellcolor{midorange} 0.355 $\pm$ 0.232
& \cellcolor{loworange} 7.523 \\

LBM, source + tumor mask 
& \cellcolor{bestgreen}\textbf{0.940 $\pm$ 1.085}
& \cellcolor{bestgreen}\textbf{0.114 $\pm$ 0.034}
& \cellcolor{bestgreen}\textbf{0.429 $\pm$ 0.185}
& \cellcolor{bestgreen}\textbf{4.716} \\

LBM, source + predicted mask 
& \cellcolor{midgreen}{0.985 $\pm$ 1.128}
& \cellcolor{midgreen}{0.115 $\pm$ 0.034}
& \cellcolor{midgreen}{0.356 $\pm$ 0.229}
& \cellcolor{midorange}{5.107} \\

LDM, source + tumor mask 
& \cellcolor{loworange} 1.122 $\pm$ 1.248
& \cellcolor{loworange} 0.136 $\pm$ 0.037
& \cellcolor{loworange} 0.322 $\pm$ 0.174
& \cellcolor{midgreen} 4.786 \\

\bottomrule
\end{tabular}
\end{table}

\begin{figure}[!ht]
\centering
\includegraphics[width=\textwidth]{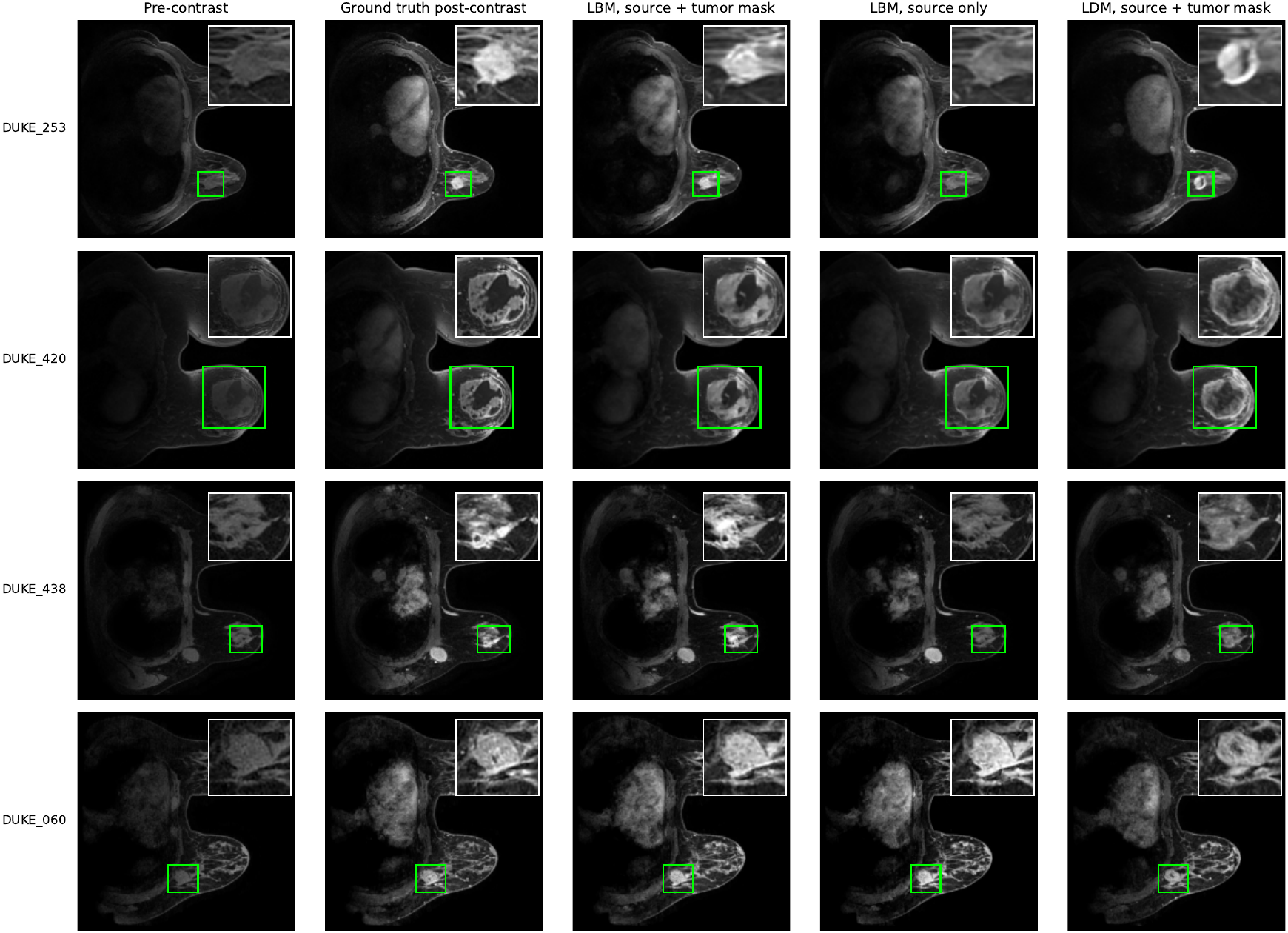}
\caption{Qualitative comparison on DUKE validation cases. From left to right, columns show the pre-contrast input, ground-truth peak-enhanced target, LBM with tumor-mask conditioning, LBM with source-only conditioning, and the evaluated LDM baseline with tumor-mask conditioning. Tumor-mask-conditioned predictions use the ground-truth validation tumor mask as conditioning input.}
\label{fig:qualitative_source_vs_mask}
\end{figure}

\section{Results}
\label{sec:results}

\textbf{Quantitative validation performance:} We evaluated the proposed models on 91 DUKE validation cases. Table~\ref{tab:validation} compares LBM with source-latent conditioning, LBM with source-latent and tumor-mask conditioning, and the evaluated LDM baseline with tumor-mask conditioning. For the tumor-mask-conditioned models, ground-truth validation tumor masks or predicted masks from a nn-Unet model trained on pre-contrast images were used as conditioning inputs. Adding tumor-mask conditioning to LBM improved all reported metrics relative to source-only conditioning, reducing MSE from 1.023 to 0.940, reducing LPIPS from 0.119 to 0.114, increasing tumor SSIM from 0.355 to 0.429, and reducing FRD from 7.523 to 4.716. The tumor-mask-conditioned LBM also outperformed the evaluated tumor-mask-conditioned LDM baseline on this internal validation. Using predicted masks slightly decreased the validation performance.  

\textbf{Qualitative comparison}: Qualitative examples are shown in Figure~\ref{fig:qualitative_source_vs_mask}. The examples compare the synthesized peak-enhanced images across conditioning strategies and illustrate the effect of tumor-mask conditioning on localized enhancement synthesis.

\section{Discussion}
\label{sec:discussion}

The proposed formulation targets the specific structure of virtual contrast synthesis. Unlike unconditional or text-to-image latent diffusion, LBM does not need to generate breast anatomy from scratch. The pre-contrast image already determines the anatomy of the patient, the shape of the breast, and much of the image texture. The remaining task is to infer contrast-dependent signal changes. A bridge formulation therefore provides a natural inductive bias: the synthesis trajectory is anchored to the observed pre-contrast image and transported toward the peak-enhanced image.

However, pre-contrast to post-contrast synthesis has an inherent information-theoretic limitation. True enhancement depends on vascularity, perfusion, permeability, contrast timing, and acquisition-specific factors that are not directly measured in the pre-contrast image. Therefore, the model cannot be expected to recover the exact patient-specific enhancement in all cases. In practice, a supervised synthesis model may learn the conditional expectation of peak enhancement given the pre-contrast appearance and conditioning inputs. This can produce a plausible average enhancement pattern, but it may fail when the true enhancement deviates from the learned population-level relationship. This distinction is important for diagnostic safety, because hallucinated or attenuated enhancement could affect lesion conspicuity, radiomic features, or downstream segmentation.

The validation comparison suggests that explicit tumor-region conditioning is useful for virtual enhancement synthesis. The segmentation-conditioned model improves tumor-region structure and radiomic realism over the source-only model while also reducing global MSE. This supports the intuition that tumor localization information can guide the latent vector field toward the region where enhancement is most clinically important. At the same time, the reported validation result uses ground-truth tumor masks, which represents an optimistic setting. In practical deployment, tumor masks would need to be predicted from the pre-contrast image, and segmentation errors could propagate into the synthesis model. The current results should therefore be interpreted as measuring the potential benefit of tumor-aware conditioning, rather than the performance of a fully automatic clinical pipeline.

Compared with previous breast MRI synthesis studies based on direct regression, U-Net translation, conditional GANs, or diffusion models, LBM provides a source-anchored transport formulation for paired synthesis. This is well matched to the MAMA-SYNTH setting, where source and target slices are spatially aligned and the desired output differs mainly by contrast-dependent enhancement. Nevertheless, the comparison in this study is limited to one evaluated LDM baseline and one internal validation cohort, so broader conclusions require additional experiments across implementations and datasets.

Several limitations remain. First, MSE is sensitive to intensity calibration and localized high-enhancement outliers, so small errors in highly enhancing tumor or vessel regions can dominate the score. Second, the VAE compression bottleneck may attenuate high-frequency enhancement patterns and extreme intensity tails. Third, the validation cohort is modest in size, and performance may be affected by domain shift across scanners, institutions, acquisition protocols, and patient populations. Fourth, the current framework uses tumor-mask conditioning; therefore, at inference time, tumor segmentation must first be obtained from the pre-contrast image, which adds an additional dependency to the pipeline. Future work should investigate DCE-MRI-specific VAE fine-tuning for preserving enhancement tails, uncertainty-aware intensity calibration, external multi-institutional validation, and approaches that remove the need for an explicit inference-time tumor mask by incorporating segmentation information only as training-time loss guidance or auxiliary supervision.

\section{Conclusion}
\label{sec:conclusion}

We presented latent bridge matching for pre- to peak-enhanced breast DCE-MRI synthesis in the MAMA-SYNTH challenge setting. By learning a conditional vector field between paired pre-contrast and peak-enhanced latents, the method provides a source-anchored formulation for patient-specific image-to-image translation. In internal validation, tumor-mask conditioning improved performance over source-only conditioning, suggesting that explicit tumor-region guidance may be beneficial for virtual contrast enhancement. However, the reported tumor-mask-conditioned results use validation tumor masks and should be interpreted as an upper-bound setting rather than a fully automatic pipeline. Further evaluation is needed with predicted tumor masks, external test cohorts, scanner- and institution-level domain shifts, and task-based clinical endpoints before assessing clinical utility.

\section*{Grants and Founding}
Authors acknowledge financial support from the European Union’s Horizon research and
innovation programme under grant agreement: CHAIMELEON n° 952172 , ImmunoSABR n°
733008, EuCanImage n° 952103, IMI-OPTIMA n° 101034347, AIDAVA (HORIZON-HLTH2021-TOOL-06) n°101057062, REALM (HORIZON-HLTH-2022-TOOL-11) n° 101095435,
RADIOVAL (HORIZON-HLTH-2021-DISEASE-04-04) n°101057699, GLIOMATCH n°
101136670 and EUCAIM (DIGITAL-2022-CLOUD-AI-02) n°101100633. The research of
H.C.W. is partially supported by the Dutch Cancer Society (KWF Kankerbestrijding)
(project no. 2021-PoC/14449).
%
%

\bibliography{bib}

@inproceedings{Osuala2024PrePostSegmentation,
  author    = {Osuala, Richard and Joshi, Smriti and Tsirikoglou, Apostolia and Garrucho, Lidia and Pinaya, Walter H. L. and Diaz, Oliver and Lekadir, Karim},
  title     = {Pre- to Post-Contrast Breast MRI Synthesis for Enhanced Tumour Segmentation},
  booktitle = {Medical Imaging 2024: Computer-Aided Diagnosis},
  volume    = {12926},
  pages     = {129260Y},
  year      = {2024},
  publisher = {SPIE},
  doi       = {10.1117/12.3006961},
  note      = {arXiv:2311.10879}
}

@inproceedings{Osuala2024ContrastKineticsLDM,
  author    = {Osuala, Richard and Lang, Daniel M. and Verma, Preeti and Joshi, Smriti and Tsirikoglou, Apostolia and Skorupko, Grzegorz and Kushibar, Kaisar and Garrucho, Lidia and Pinaya, Walter H. L. and Diaz, Oliver and Schnabel, Julia A. and Lekadir, Karim},
  title     = {Towards Learning Contrast Kinetics with Multi-Condition Latent Diffusion Models},
  booktitle = {Medical Image Computing and Computer Assisted Intervention -- MICCAI 2024},
  series    = {Lecture Notes in Computer Science},
  volume    = {15005},
  pages     = {713--723},
  year      = {2024},
  publisher = {Springer Nature Switzerland},
  doi       = {10.1007/978-3-031-72086-4_67},
  note      = {arXiv:2403.13890}
}

@article{Osuala2025DynamicTumorCGAN,
  author  = {Osuala, Richard and Joshi, Smriti and Tsirikoglou, Apostolia and Garrucho, Lidia and Pinaya, Walter H. L. and Lang, Daniel M. and Schnabel, Julia A. and Diaz, Oliver and Lekadir, Karim},
  title   = {Simulating Dynamic Tumor Contrast Enhancement in Breast MRI Using Conditional Generative Adversarial Networks},
  journal = {Journal of Medical Imaging},
  volume  = {12},
  number  = {S2},
  pages   = {S22014},
  year    = {2025},
  doi     = {10.1117/1.JMI.12.S2.S22014},
  note    = {arXiv:2409.18872}
}

@article{Iyad2023GadoliniumReview,
  author  = {Iyad, N. and Ahmad, M. S. and Alkhatib, S. G. and Hjouj, M.},
  title   = {Gadolinium contrast agents: challenges and opportunities of a multidisciplinary approach: Literature review},
  journal = {European Journal of Radiology Open},
  volume  = {11},
  pages   = {100503},
  year    = {2023},
  doi     = {10.1016/j.ejro.2023.100503}
}

@misc{chadebec2025lbmlatentbridgematching,
      title={LBM: Latent Bridge Matching for Fast Image-to-Image Translation}, 
      author={Clément Chadebec and Onur Tasar and Sanjeev Sreetharan and Benjamin Aubin},
      year={2025},
      eprint={2503.07535},
      archivePrefix={arXiv},
      primaryClass={cs.CV},
      url={https://arxiv.org/abs/2503.07535}, 
}

@inproceedings{Rombach2022LatentDiffusion,
  author    = {Rombach, Robin and Blattmann, Andreas and Lorenz, Dominik and Esser, Patrick and Ommer, Bjorn},
  title     = {High-Resolution Image Synthesis with Latent Diffusion Models},
  booktitle = {Proceedings of the IEEE/CVF Conference on Computer Vision and Pattern Recognition},
  pages     = {10684--10695},
  year      = {2022},
  url       = {https://arxiv.org/abs/2112.10752}
}

@inproceedings{Lipman2023FlowMatching,
  author    = {Lipman, Yaron and Chen, Ricky T. Q. and Ben-Hamu, Heli and Nickel, Maximilian and Le, Matthew},
  title     = {Flow Matching for Generative Modeling},
  booktitle = {International Conference on Learning Representations},
  year      = {2023},
  url       = {https://openreview.net/forum?id=PqvMRDCJT9t}
}

@inproceedings{Liu2023RectifiedFlow,
  author    = {Liu, Xingchao and Gong, Chengyue and Liu, Qiang},
  title     = {Flow Straight and Fast: Learning to Generate and Transfer Data with Rectified Flow},
  booktitle = {International Conference on Learning Representations},
  year      = {2023},
  url       = {https://openreview.net/forum?id=XVjTT1nw5z}
}

@article{Garrucho2025MAMAMIA,
  author  = {Garrucho, L. and Kushibar, K. and Reidel, C. A. and others},
  title   = {A large-scale multicenter breast cancer {DCE-MRI} benchmark dataset with expert segmentations},
  journal = {Scientific Data},
  volume  = {12},
  pages   = {453},
  year    = {2025},
  doi     = {10.1038/s41597-025-04707-4}
}

@misc{MAMASYNTH2026,
  author       = {{MAMA-SYNTH Challenge Organizers}},
  title        = {{MAMA-SYNTH}: Breast {DCE-MRI} Synthesis Challenge},
  year         = {2026},
  howpublished = {Official challenge website and GitHub repository},
  url          = {https://www.ub.edu/mama-synth/mama-synth},
  note         = {Accessed 2026-07-13}
}

\end{document}